\documentclass[aps,twocolumn,showpacs,preprintnumbers,prl]{revtex4}
\usepackage{graphicx}
\usepackage{epsfig,latexsym}

\newcommand{\be}{\begin{equation}}
\newcommand{\ee}{\end{equation}}
\newcommand{\ber}{\begin{eqnarray}}
\newcommand{\eer}{\end{eqnarray}}

\begin{document}
\title{The Sound of the Little Bangs%\\ or\\
%Acoustic(s) (Spectra) of Heavy Ion Collisions
}

\author{\'Agnes M\'ocsy \email{mocsy@bnl.gov}} \affiliation{ Pratt Institute, Department of Math and Science, Brooklyn, NY 11205, USA\\  Frankfurt Institute for Advanced Studies, Ruth-Moufang-Str.1, D-60438 Frankfurt am Main, Germany}
\author{Paul Sorensen\email{prsorensen@bnl.gov}} \affiliation{
Physics Department, Brookhaven National Laboratory, Upton NY 11973, USA }

\begin{abstract}
  Data from ultrarelativistic heavy-ions collisions show evidence for
  temperature-fluctuations on the freeze-out surface of the expanding
  fireball. These may be remnants of density inhomogeneities in the
  initial collision overlap region. We present a power-spectrum
  analysis for heavy-ion collisions analogous to the analysis of the
  cosmic microwave background radiation. We use a Glauber model for
  eccentricity to extract the transfer-function needed to produce the
  observed spectrum and discuss its relation to the mean-free-path of
  the matter created in the collisions.
\end{abstract}
%\pacs{11.15.Ha, 12.38.Aw}
%\preprint{BNL-NT  07/26 }
\maketitle

A commonly quoted goal of the heavy-ion programs at Brookhaven Lab
(BNL) and CERN, is to recreate conditions similar to those shortly
after the Big Bang when the universe was filled with a quark-gluon
plasma (QGP). QGP can be created by smashing heavy nuclei together at
relativistic speeds in collisions called "little bangs". In the past
few decades many theoretical and experimental advances have been made
in the study of heavy-ion collisions. Comparisons to the early
universe, however, have been scarce \cite{hatsuda,mishra}. In this
paper we explore an analogy between heavy-ion collisions and Big Bang
cosmology. Using the heavy-ion equivalent of the map of the cosmic
microwave background radiation (CMB), we determine the power-spectrum
for the little bangs and we estimate the \textit{transfer-function}
necessary to produce the spectrum from the initial
conditions.

The hot fireball created in little bangs rapidly expands and cools and
when cold enough forms hadrons. Eventually, the system spreads out
enough that hadrons stop interacting. This is called the surface of
last scattering or freeze-out. The particles then free-stream to the
detectors. From measurements of the number, mass, and momentum of
these particles, we must infer the properties of the matter that
emitted them and extract the essential physics of the QGP. One of the
most important discoveries at the Relativistic Heavy Ion Collider
(RHIC) at BNL is that the tiny speck of QGP matter produced in the
little bangs behaves much like a liquid \cite{liquid}. This finding is
based on the observation that the spatial asymmetries in the initial
overlap zone show up as asymmetries in the momentum distributions of
final state particles. The observed anisotropy is typically
represented by the second Fourier component ($v_2$) of the azimuthal
distribution of observed particles relative to the
reaction-plane~\cite{noteplane}. $v_2$ most strongly reflects the
almond shape of the initial nuclear overlap region for non central
collisions. The magnitude of $v_2$ can be described surprisingly well
with ideal relativistic hydrodynamic models suggesting a liquid-like
behavior with minimal viscosity. The success of these models seems to
indicate that the mean-free-path of interactions for the systems
constituents is significantly smaller than the size of the system.

Experiments at RHIC have also discovered correlations between
particles that extend over a broad range in the longitudinal direction
but are narrow in the azimuthal (transverse) direction forming a
ridge~\cite{data}. A number of different scenarios have been proposed
to explain the ridge
\cite{trainor-jet,majumder,shuryak,gavin,voloshin,voloshin-jet,mclerran,sorensen1}.
These include minimum-bias (soft) jets in Au+Au collisions
\cite{trainor-jet}, soft gluons radiated by hard partons traversing
the overlap region \cite{majumder}, beam-jets boosted by the radial
expansion \cite{shuryak}, viscous broadening \cite{gavin}, and
flux-tube like structures boosted by the radial expansion
\cite{voloshin-jet, mclerran}. The extent of the correlation in the
longitudinal direction requires by causality that it must be
established very early in the collisions~\cite{mclerran}.  One of us
(PS) proposed that the correlation structures may be
understood in terms of fluctuations of higher Fourier components of
$v_n$, particularly $\sqrt{\langle v_3^2\rangle}$, that arise from
anisotropies in the initial energy density converted into momentum
space during the expansion~\cite{sorensen1}.  It was subsequently shown with the
NEXSPHERIO hydrodynamic model, that indeed, lumpy initial conditions
lead to structures similar to those observed in the two-particle
correlation measurements \cite{brazil}. Alver and Roland \cite{v3}
used the RQMD model to show that lumpiness in the initial collision
geometry can lead to a $\sqrt{\langle v_3^2\rangle}$ in the azimuthal
particle production. Petersen \textit{et al.} \cite{hannah} carried
out a similar analysis using an event-by-event hydrodynamic model.

An analogy between the expansion of heavy-ion collisions starting from
a lumpy initial energy density and the expansion of the universe
starting with quantum fluctuations stretched to cosmological sizes was
first pointed out by Mishra \textit{et al.}~\cite{mishra}. They also
proposed that the RMS values of $v_n$ ($\sqrt{\langle v_n^2\rangle}$)
could be measured in heavy ion collisions analogous to the
power-spectrum extracted from the CMB.  They didn't however make a
connection between $\sqrt{\langle v_n^2\rangle}$ and the already
existing two-particle correlation measurements.  In this work we use
transverse momentum ($p_T$) correlations published by the STAR
collaboration \cite{star} to extract the power-spectrum for heavy-ion
collisions. Since the $p_T$ spectra reflects the temperature, $p_T$
correlations are sensitive to local temperature fluctuations. These
measurements are directly analogous therefore, to the maps of the
CMB. We use a Monte Carlo Glauber model~\cite{mcg} for initial
eccentricities to extract the transfer-function required to convert
the initial coordinate-space anisotropy into the anisotropy seen in
momentum-space.  This analysis facilitates a more direct comparison
between relativistic heavy-ion collisions and the early universe.

\begin{figure}[th]
\includegraphics[width=9cm]{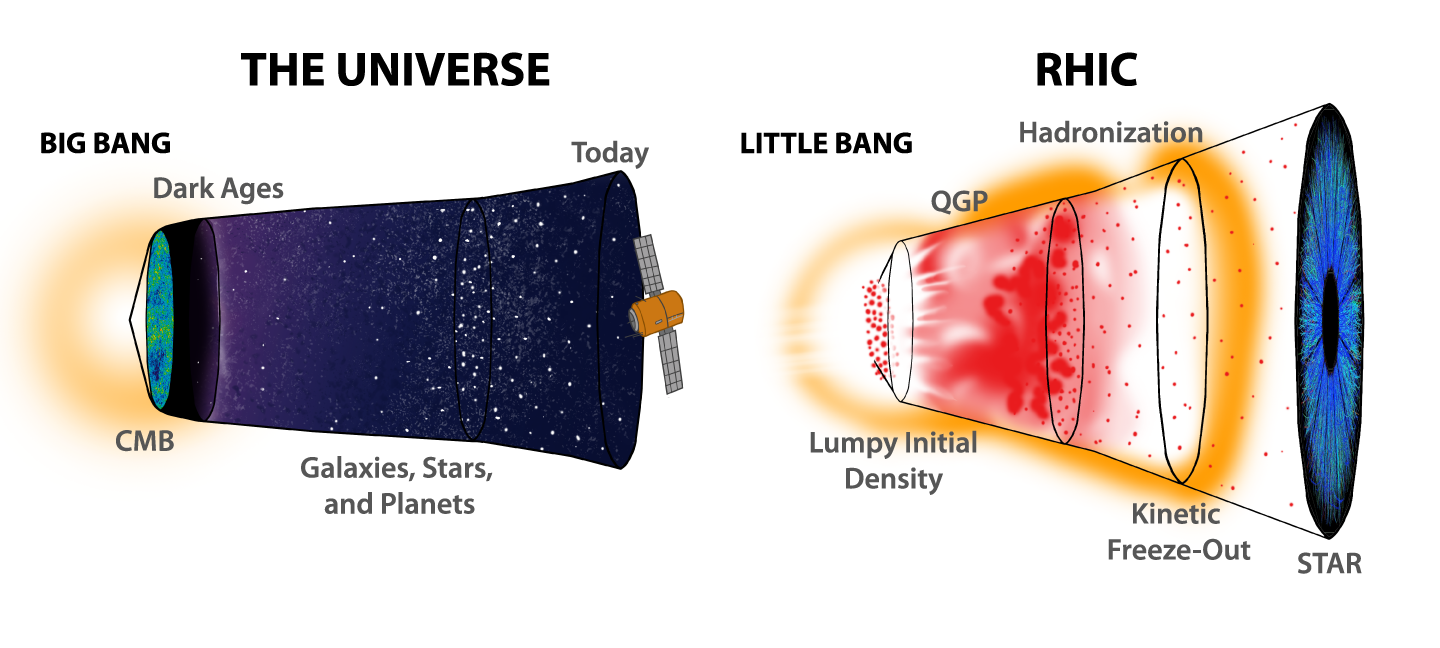}
\caption{ Schematic of the expansion of the universe after the Big
  Bang (left) and the expansion of a fireball after little bangs (right).
  The illustration is by Alexander Doig.}
\label{fig:expansion}
\end{figure}

\textit{Analogy with Big Bang Cosmology:} Measurements of the CMB
reveal temperature fluctuations corresponding to over- or
under-densities present at the surface of last scattering at about
300,000 years after the Big Bang~\cite{cmb}.  These density
fluctuations ultimately explain the structure in our universe
(Fig. \ref{fig:expansion} left). Just as quantum fluctuations
stretched to cosmic sizes by inflation show up in the CMB, we expect
fluctuations from the beginning of the little bangs to show up in
heavy-ion data (Fig. \ref{fig:expansion} right).  Measuring
temperature-fluctuations in the CMB required precise measurements at
more than two million points in the sky.  Enough photons are detected
at each point to reconstruct the black-body spectrum from which the
temperature is determined. In a heavy-ion collision, a few thousand
particles are created at most, so a similar map cannot be made for
each collision. But whereas we only observe one universe, billions of
collisions are created in the lab.  By studying $p_T$ correlation data
(sensitive to local changes in the $p_T$-spectra and thus the
temperature) accumulated from millions of these collisions, we can
search for evidence of hotspots on the surface of last scattering.

\begin{figure}[thb]
\includegraphics[width=4cm]{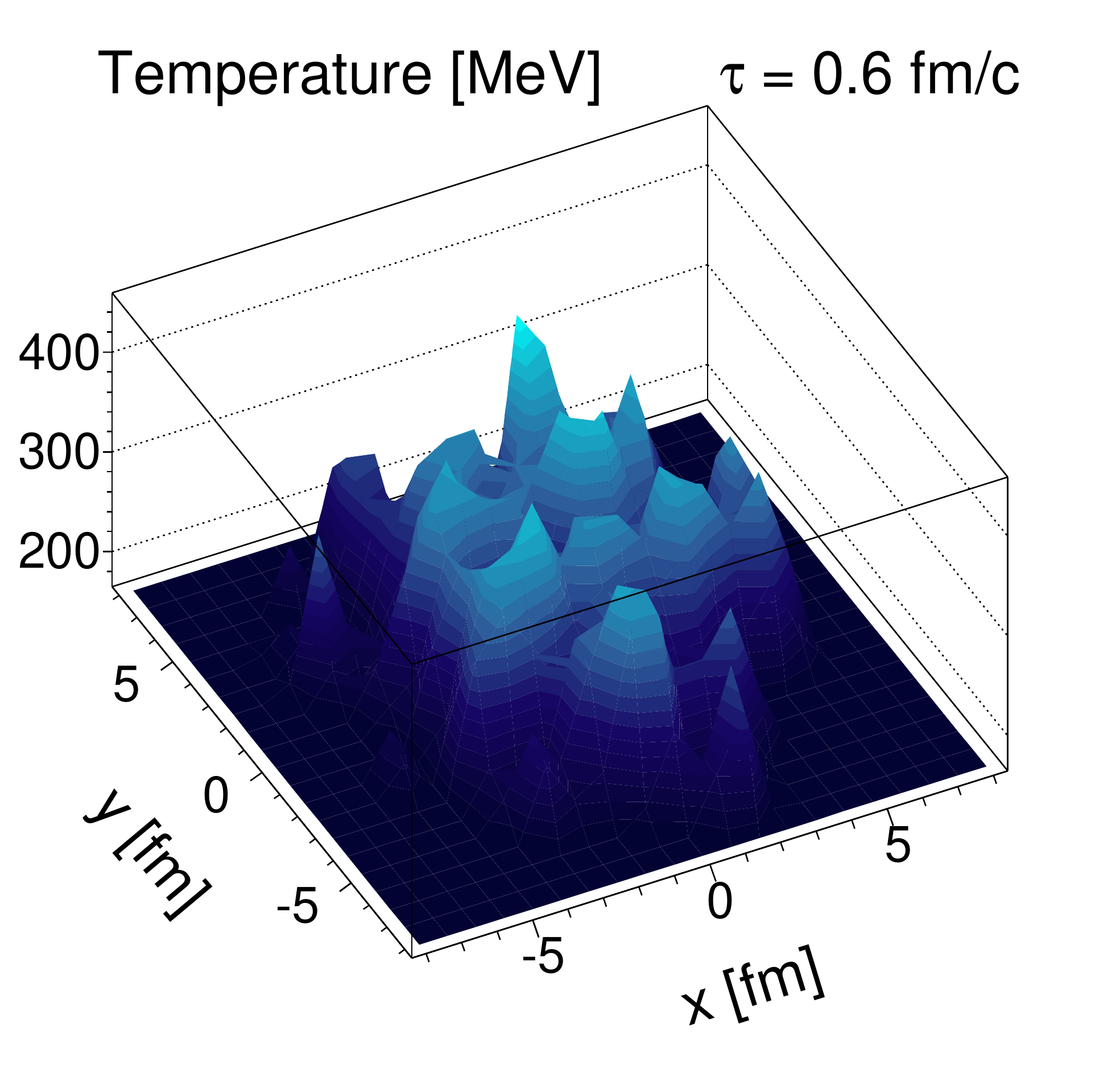}
\includegraphics[width=4cm]{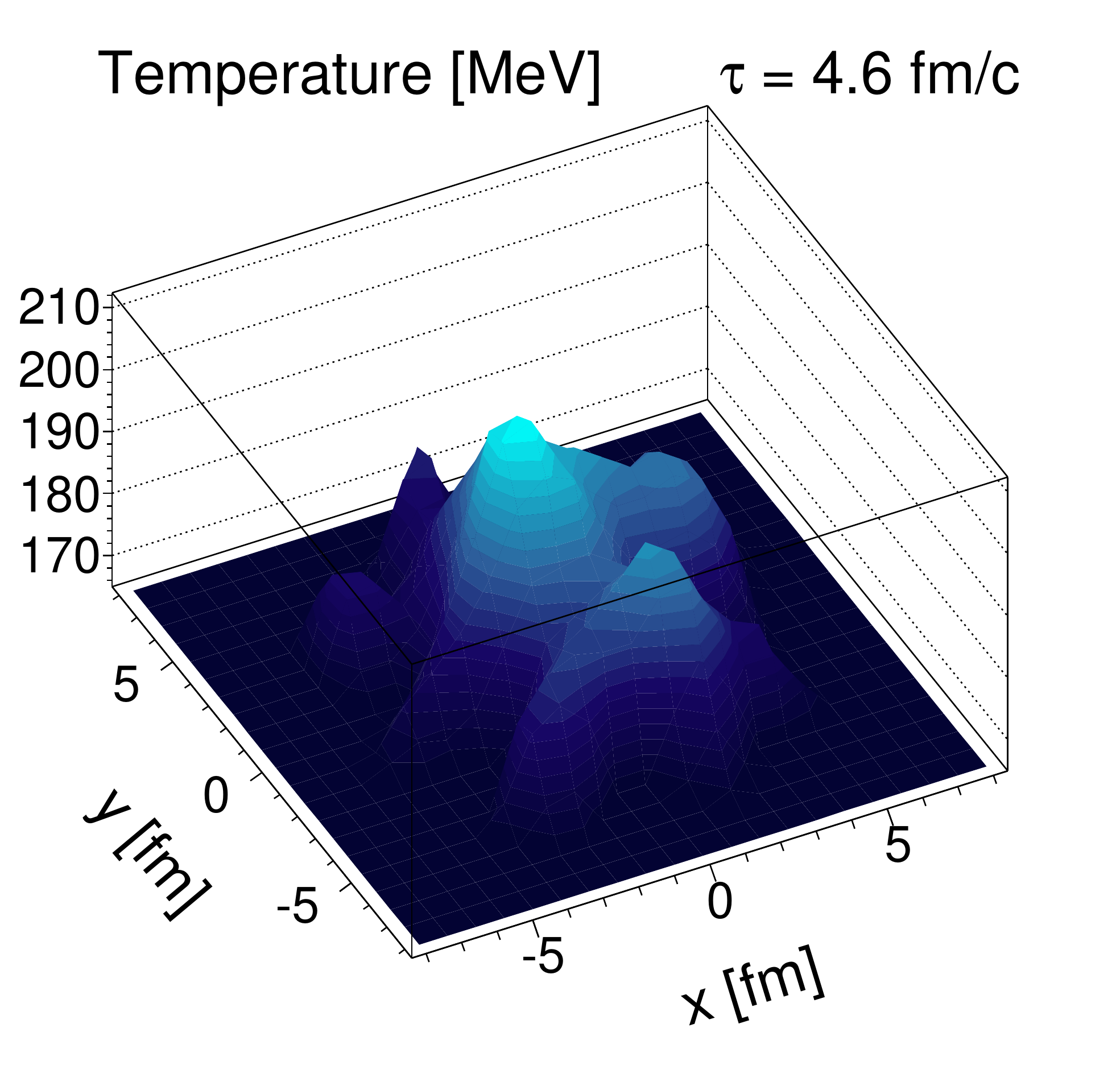}
\caption{Temperature profile in the transverse plane for mid-rapidity at proper time $\tau=0.6$ (left) and 4.6 fm/c (right). }
\label{fig:hotspots}
\end{figure}

\textit{Survival of Density Fluctuations and Various Scales:} In
Fig.~\ref{fig:hotspots} we show the temperature fluctuations calculated
from a heavy-ion event generator near the beginning of the expansion
and 4 $fm/c$ later. We determined these temperature profiles in the
transverse plane at mid-rapidity by translating the energy-density
profiles of Werner \textit{et al.} \cite{werner} into temperature,
using the parametrized lattice QCD results for energy-density vs
temperature from~\cite{lattice}. The simulations indicate that
collisions of Au nuclei (12~fm across), may contain hotspots of size
$l_{spot} \approx 1.5$ fm and that remnants of those hotspots
persists during the collisions evolution. We also consider the lengths
of the acoustic horizon $H$ and the mean-free-path $l_{mfp}$ of the
systems constituents to be important.

\begin{figure}[htb]
\includegraphics[width=5cm]{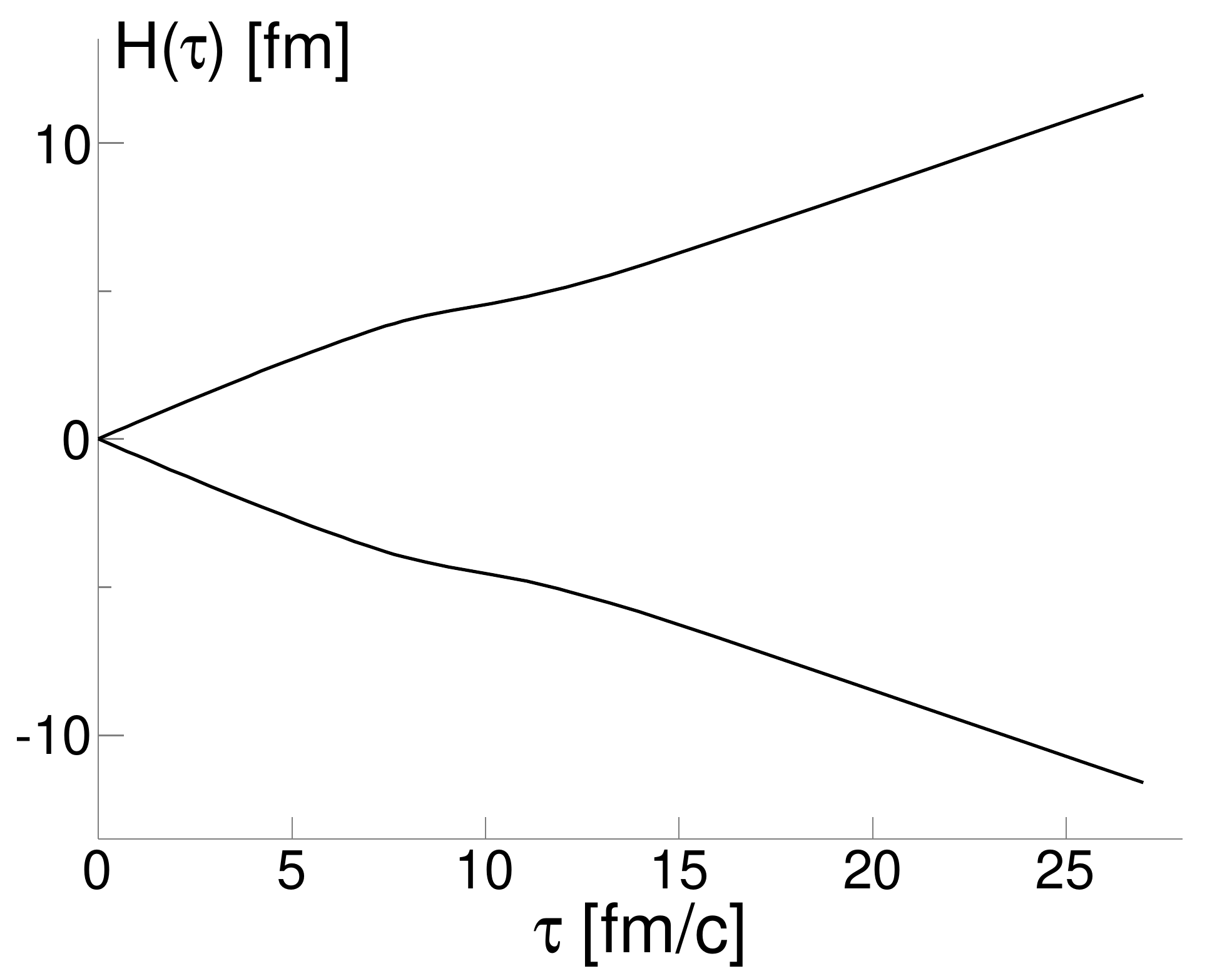}
\caption{The acoustic horizon for heavy ion collisions at RHIC.}
\label{fig:horizon}
\end{figure}

The acoustic horizon defines how far mechanical information can have
propagated through the medium at time $\tau$:
$H(\tau)=\int_{0}^{\tau_{fo}} c_s(\tau)d\tau$, where $\tau_{fo}$ is
the freeze-out time. This relates to the growth rate of $l_{spot}$. We
determined $H(\tau)$ from lattice data on the speed of sound ($c_s$)
vs energy density~\cite{pasi} and a hydrodynamic model to specify
the energy density vs $\tau$~\cite{hydro}. Fig.~\ref{fig:horizon}
shows the acoustic horizon for QCD matter. The phase-transition from
QGP to hadron-gas can be seen as a flattening in the slope of $H$ at $\tau
\approx 10$~$fm/c$ when $H$ is about 5 $fm$.  The acoustic horizon
also dominates the time dependence of the sound that an observer
inside the medium would hear (see \cite{soundweb} and
\cite{sound}). The sound at freeze-out is composed of a superposition
of different waves with different frequencies that can be determined
from the two-particle momentum correlation data.  The horizon defines
when frequencies can be heard: Only after half a wavelength fits
inside the horizon would that wavelength become "audible". This is the
same effect that leads to the lack of large scale fluctuations in the
CMB.

The fact that hydrodynamic models do a reasonable job of predicting
the value of $v_2$ suggests that $l_{mfp}$ can be considered small
compared to the size of the system. By examining the power-spectrum of
heavy-ion collisions which includes information for all values of $n$
(beyond just $n=2$ or $n=3$), we hope to better constrain
$l_{mfp}$. As we increase $n$, we reduce the length scale probed. We
only expect an efficient conversion of coordinate-space anisotropies
into momentum space when $l_{mfp}\ll2\pi\langle R \rangle/n$ where
$\langle R \rangle$ is the average radius of the systems
constituents~\cite{noten1}.

\textit{Power-Spectrum and Transfer-Function:} We determine the
power-spectrum from two-particle momentum correlations (related to
$(p_{t1}-\langle p_t\rangle)(p_{t2}-\langle p_t\rangle)$) vs the
relative azimuthal angles between the particles $\Delta\phi$
\cite{star}.  A narrow peak positioned around small angle separation
is observed in the data. This tells us that if a particle comes out
with above average $p_T$, then the nearby particles also tend to have
large $p_T$. This is consistent with expectations from hotspots on the
surface of last scattering. The correlation of these fast particles
suggests that they are born out of the same high-density,
high-temperature lump. We will not attempt to decompose the
correlation into different components, i.e. jets and resonances and
background. The power-spectrum we extract from data should and does
contain all these contributions.  We argue that jets do not dominate
the observed correlation because the correlations are too large in
magnitude, too narrow in $\phi$, and too broad in $\eta$: there will
likely be some contribution but it is suppressed by
~1/multiplicity). As for resonances, if a hotspot is there, it will
emit massive and/or high momentum particles. The decay of a hotspot
can proceed through decays into resonances. The power-spectrum
reflects all these contributions. Our interpretation of the
correlations data in terms of hotspots is supported by several pieces
of ancillary evidence. 1) $v_n$-fluctuations are close to what we
expect from density fluctuations from several models and the
two-particle correlations data also match what we expect from these
models \cite{sorensen2,brazil,v3,hannah}. This gives us confidence
that the correlations are dominated by the over- and under-densities
at the start of the expansion phase.  2) An improved description of
particle $p_T$ spectra is obtained when temperature fluctuations are
considered \cite{zhangbu}.

\begin{figure}[htb]
\includegraphics[width=4.25cm]{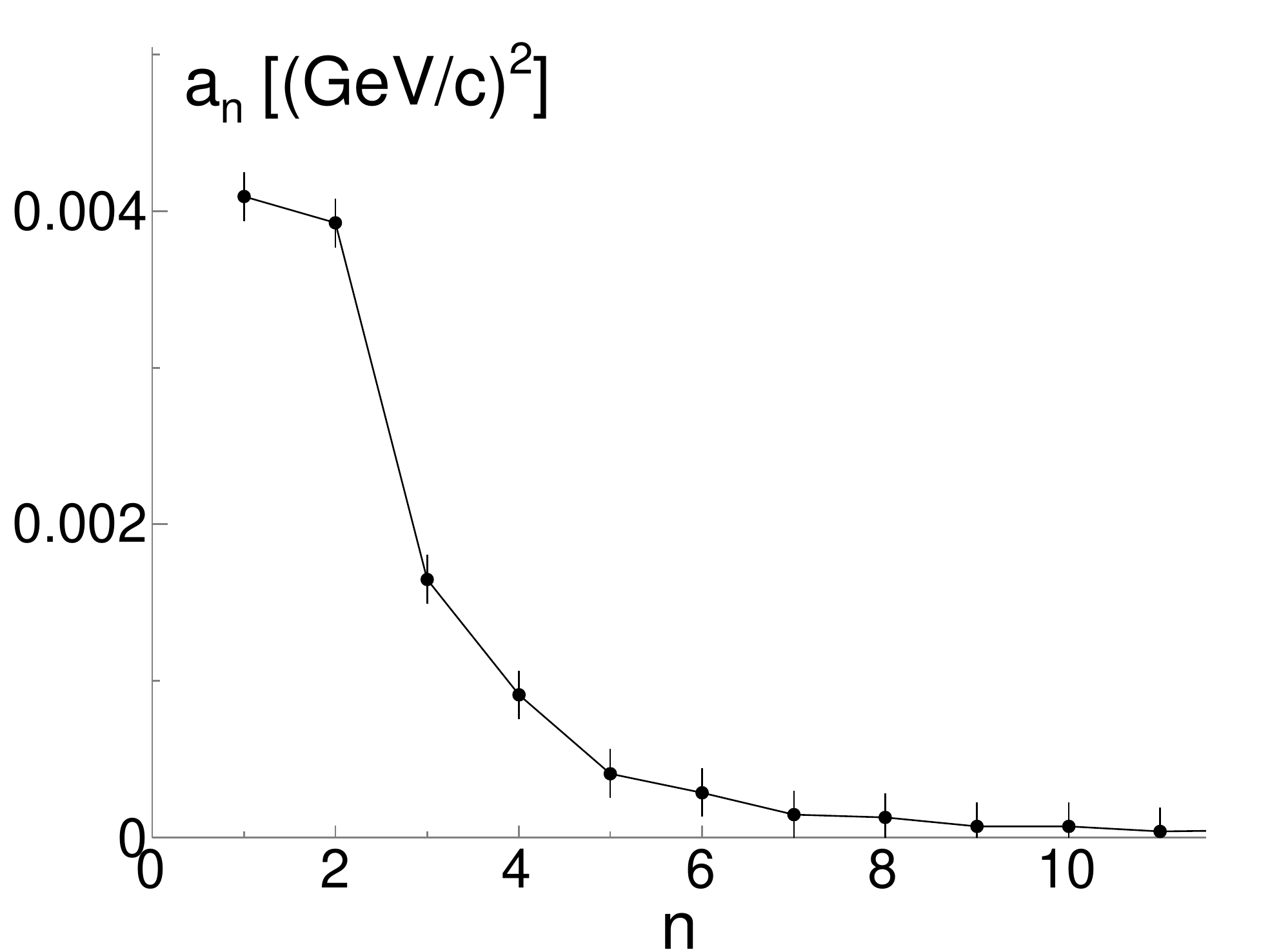}
\includegraphics[width=4.25cm]{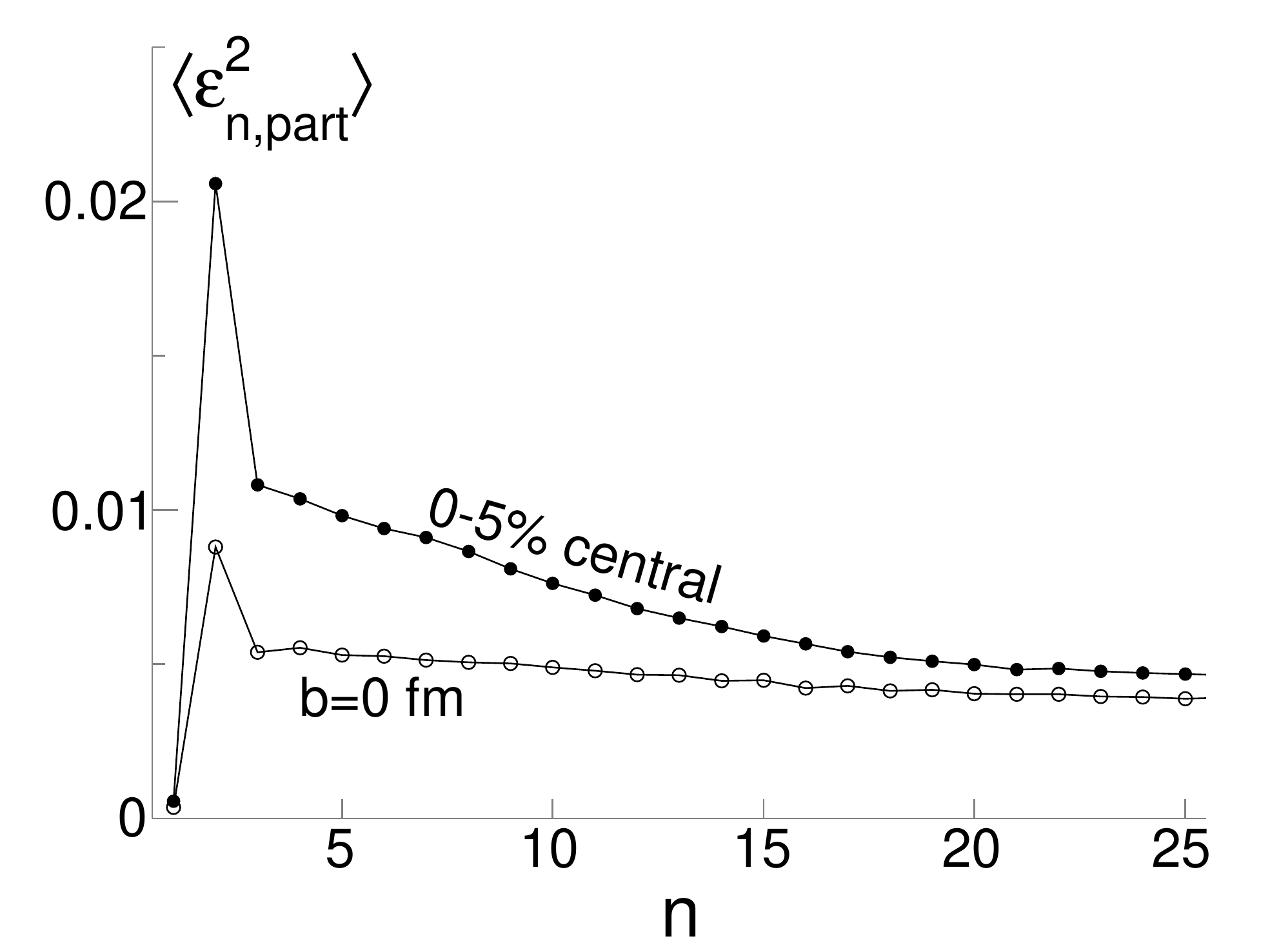}
\caption{Left: The power-spectrum versus harmonic number at central
  rapidity. Right: Participant eccentricity for arbitrary harmonic n
  for either perfectly central Au+Au collisions or the 5\% most
  central. The intrinsic deformation of the Au nucleus
  is included.}
\label{fig:PS}
\end{figure}

The $p_T$ correlations vs relative angles between the emitted
particles ($\Delta\phi$ and $\Delta\eta$) where parametrized in the
STAR paper~\cite{star}.  To extract the power-spectrum, we use that
parametrization with $\Delta\eta=0$ and Fourier-transform the
correlation function versus $\Delta\phi$. The coefficients 
\be
a_n=\frac{2}{\pi}\int_0^\infty f(\Delta\phi)\cos{(n\Delta\phi)}
d(\Delta\phi) .  
\ee
vs harmonic $n$ make up the power-spectrum. If we had used number
correlations instead of $p_T$ correlations $a_{n}\approx
v_{n}^{2}$~\cite{trainor}. The power-spectrum for little bangs is
shown in Figure \ref{fig:PS} (left).

Having extracted the power-spectrum from the azimuthal correlations of
observed particles, we can compare that to the azimuthal distribution
of matter in the initial overlap zone. This is calculated using the
participant eccentricity ($\varepsilon_{n,part}$) for all harmonics $n$
as in Ref.~\cite{v3}. The fluctuations in the initial geometry cause
the major axis of the eccentricity to fluctuate away from the
reaction-plane direction. $\varepsilon_{n,part}$ is the eccentricity
calculated along the major-axis. Fig.~\ref{fig:PS} (right) shows
$\langle\varepsilon_{n,part}^2 \rangle$ from a Monte Carlo Glauber
model~\cite{mcg} for perfectly central collision (b=0 fm) and for the
5\% most central collisions. The large $n=2$ term persists even for
central collisions because we include the intrinsic deformation of
the Au nucleus in our Monte Carlo. The $n=1$ term is small because the
participants are re-centered so that $\langle x\rangle=\langle
y\rangle = 0$. For $b=0$ collisions we note that the eccentricity is
nearly independent of $n$ for $n>2$. This is because for symmetric
collisions and point-like participants, $\varepsilon_{part}$ depends
only on the number of participants, independent of $n$~\cite{notesize}.
In this case, if all harmonics were converted into momentum space
equally well, the final correlation function tend to a Dirac
delta function at $\Delta\phi=0$. We expect however, that the
conversion of higher harmonic eccentricity will be damped due to the
existence of the length scale $l_{mfp}$. The conversion will
be efficient only when $l_{mfp}<2\pi\langle R \rangle/n$. We can
investigate the damping of the higher modes by plotting the
transfer-function which is the ratio of the power-spectrum in
Fig.~\ref{fig:PS} (left) to $\langle\varepsilon_{part,n}^2\rangle$ in
Fig.~\ref{fig:PS} (right).

\begin{figure}[htb]
\includegraphics[width=6.5cm]{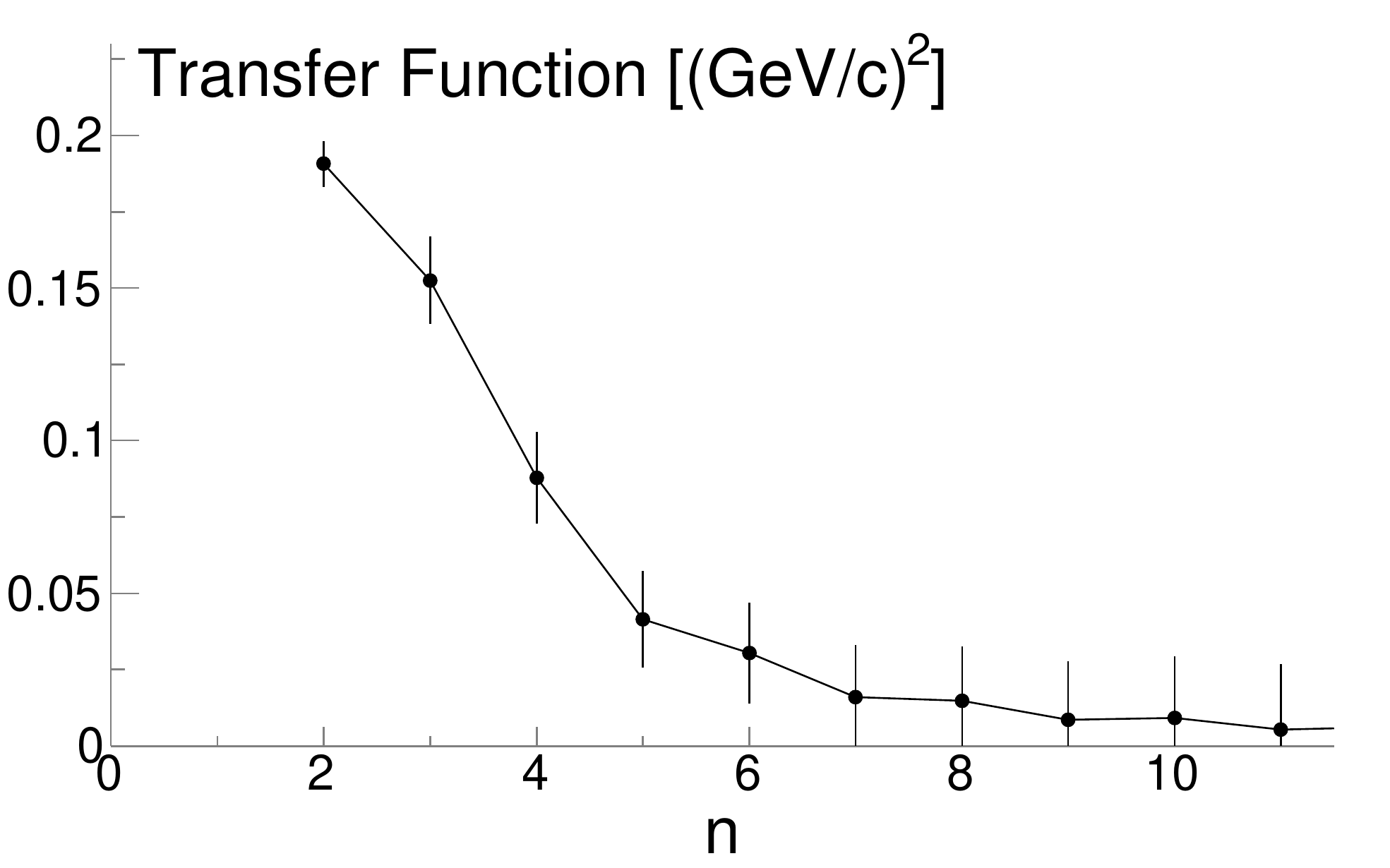}
\caption{The transfer-function for central Au+Au collisions extracted
  from $p_T$ correlations for $n>1$.}
\label{fig:transfer}
\end{figure}
Fig.~\ref{fig:transfer} shows the transfer-function
$a_n/\langle\varepsilon_{part,n}^2\rangle$.  We use
$\langle\varepsilon_{part,n}^2\rangle$ for a centrality range that
matches the data.  Understanding the shape of the transfer-function
implies that one can understand the shape of the correlations data.
The transfer-function shows that as $n$ increases, the efficiency for
converting coordinate-space anisotropy into momentum space quickly
drops off as expected from the condition that the transfer-function
should go to zero when $l_{mfp}\sim 2\pi\langle R \rangle/n$. We can make
a crude estimate for $l_{mfp}$ based on the transfer-function. If we
take $\langle R \rangle=3$ fm for the average radial position of a
participant, and $n=5$ as the harmonic beyond which the conversion is
inefficient, then we get $l_{mpf}\approx 3.5$~fm. This estimate
corresponds to a viscosity five times larger than estimates based on
the centrality dependence of $v_2$~\cite{knudsenfit}. Our crude
estimate, however, is geometry based, not accounting for the various
phases of the expansion. The authors of Ref.~\cite{gavinaziz} estimate
the viscosity from the longitudinal width of the near-side peak in
$p_T$ correlations, while our estimate of $l_{mfp}$ is based on the
transverse width.  The transfer-function should be compared to a more
complete model of heavy-ion collisions in order to understand the
effects of the expansion velocity, viscosity in the QGP, and viscosity
in the hadronic phase.

In this letter we considered the acoustics of heavy ion collisions and
discussed the analogy with the early universe. We presented the
power-spectrum from heavy ion collisions and derived the
transfer-function needed to convert spatial correlations from the
initial conditions into the $p_T$ correlations measured at RHIC. We
find the transfer-function required to describe the RHIC data can be
easily understood in terms of an inefficiency in conversion of
coordinate space anisotropy into momentum space when
$l_{mfp}>2\pi\langle R\rangle/n$. We used this transfer-function to
make a rough estimate of the mean-free-path of the systems
constituents. This approach represents a new method for determining
the characteristics of heavy-ion collisions and the QGP and should be
further investigated.

{\bf Acknowledgments:} While preparing this draft, a paper on a
similar topic was posted was posted~\cite{shuryak2}. We thank Marcus
Bleicher and Klaus Werner for providing the data for the energy
density profile and Alex Doig for preparing several illustrations. AM
thanks FIAS for the kind hospitality.

%****************************


\begin{thebibliography}{99}
\bibitem{noteplane} The reaction-plane is defined by the beam axis and
  the vector connecting the centers of the colliding nuclei.
\bibitem{notesize} Giving the participants a size causes
  $\langle\varepsilon_{n,part}^2\rangle$ to drop as a function of $n$.
  The drop occurs at $n\approx\pi\langle R \rangle/r_{part}$ where
  $r_{part}$ is the radial size of the participant and $\langle R
  \rangle \approx 3$~fm is the average radial position of a
  participant. This drop is similar to the drop we expect in the
  transfer-function due to $l_{mfp}$ and the effects could be confused
  in the data.
\end{thebibliography}
\end{document}